\documentclass[prl,showpacs,twocolumn,amsmath,amssymb,floatfix]{revtex4-1}

\usepackage{graphicx}
\usepackage{bm}
\usepackage{color}
\usepackage[applemac]{inputenc}
\usepackage{pdfpages}
 
\newcommand{\sect}[1]{\emph{#1.---}\ignorespaces} 
 \newcommand{\ndsds}{$\mathit{N^{d}S^{d}S}$}
 \newcommand{\nsds}{$\mathit{NS^{d}S}$}

\begin{document}    
  
\title{Transport spectroscopy of NS nanowire junctions with Majorana fermions}
\author{Elsa Prada$^1$, Pablo San-Jose$^2$, Ram\'on Aguado$^1$}
\affiliation{$^1$Instituto de Ciencia de Materiales de Madrid (ICMM-CSIC), Cantoblanco, 28049 Madrid, Spain\\$^2$Instituto de Estructura de la Materia (IEM-CSIC), Serrano 123, 28006 Madrid, Spain}

\date{\today} 

\begin{abstract}
We investigate transport though normal-superconductor nanowire junctions in the presence of spin-orbit coupling and magnetic field. As the Zeeman field crosses the critical bulk value $B_c$ of the topological transition, a Majorana bound state (MBS) is formed, giving rise to a sharp zero-bias anomaly (ZBA) in the tunneling differential conductance. We identify novel features beyond this picture in wires with inhomogeneous depletion, like the appearance of two MBSs inside a long depleted region for $B<B_c$. The resulting ZBA is in most cases weakly split and may coexist with Andreev bound states near zero energy. The ZBA may appear without evidence of a topological gap closing. This latter aspect is more evident in the multiband case and stems from a smooth pinch-off barrier. Most of these features are in qualitative agreement with recent experiments [Mourik {\it et al}, Science, {\bf 336}, 1003 (2012)]. We also discuss the rich phenomenology of the problem in other regimes which remain experimentally unexplored.
\end{abstract}

\maketitle
Following early ideas based on exotic p-wave superconductors~\cite{Volovik:JL99,Read:PRB00}, it has been recently predicted that Majorana quasiparticles should appear in topological insulators~\cite{Fu:PRL08} and semiconductors with strong spin-orbit (SO) coupling~\cite{Sato:PRL09,Sau:PRL10,Alicea:PRB10,Lutchyn:PRL10,Oreg:PRL10}. In proximity to s-wave superconductors, these systems behave as topological superconductors (TS) when the excitation gap is closed and reopened again:
as the gap crosses zero, Majorana bound states (MBSs) appear wherever the system interfaces with a non-topological {insulator} (see Refs. \cite{Beenakker:11,Alicea:RPP12} for reviews). 

The TS transition occurs when an external Zeeman field $B$ exceeds a critical value $B_c\equiv\sqrt{\mu^2+\Delta^2}$ defined in terms of the Fermi energy $\mu$ and the induced s-wave pairing $\Delta$ ~\cite{Lutchyn:PRL10,Oreg:PRL10}. This prediction 
has spurred a great deal of experimental activity towards detecting MBSs in hybrid superconductor-semiconductor systems. Indeed, signatures of Majorana detection have been recently reported in Ref. \cite{Delft-exp}. These experiments (and subsequently Refs. \cite{Deng:12,Das:12})
clearly show the emergence of a zero-bias anomaly (ZBA) in differential conductance $dI/dV$ measurements as $B$ increases. It has been predicted that such ZBA reflects tunneling into the MBS \cite{Sengupta:PRB01,Bolech:PRL07,Law:PRL09}. Crucially, the emerging ZBA, which signals the TS transition, should be accompanied by a closing and reopening of the excitation gap \cite{Sau:PRB10,Stanescu:PRB11}, something which is however \emph{not} observed. Other experimental findings, such as ZBA splitting and coexistence of Andreev bound states (ABSs) and MBSs \cite{Delft-exp}, need also further analysis.

\begin{figure}[t] %  figure placement: here, top, bottom, or page
   \centering
   \includegraphics[width=\columnwidth]{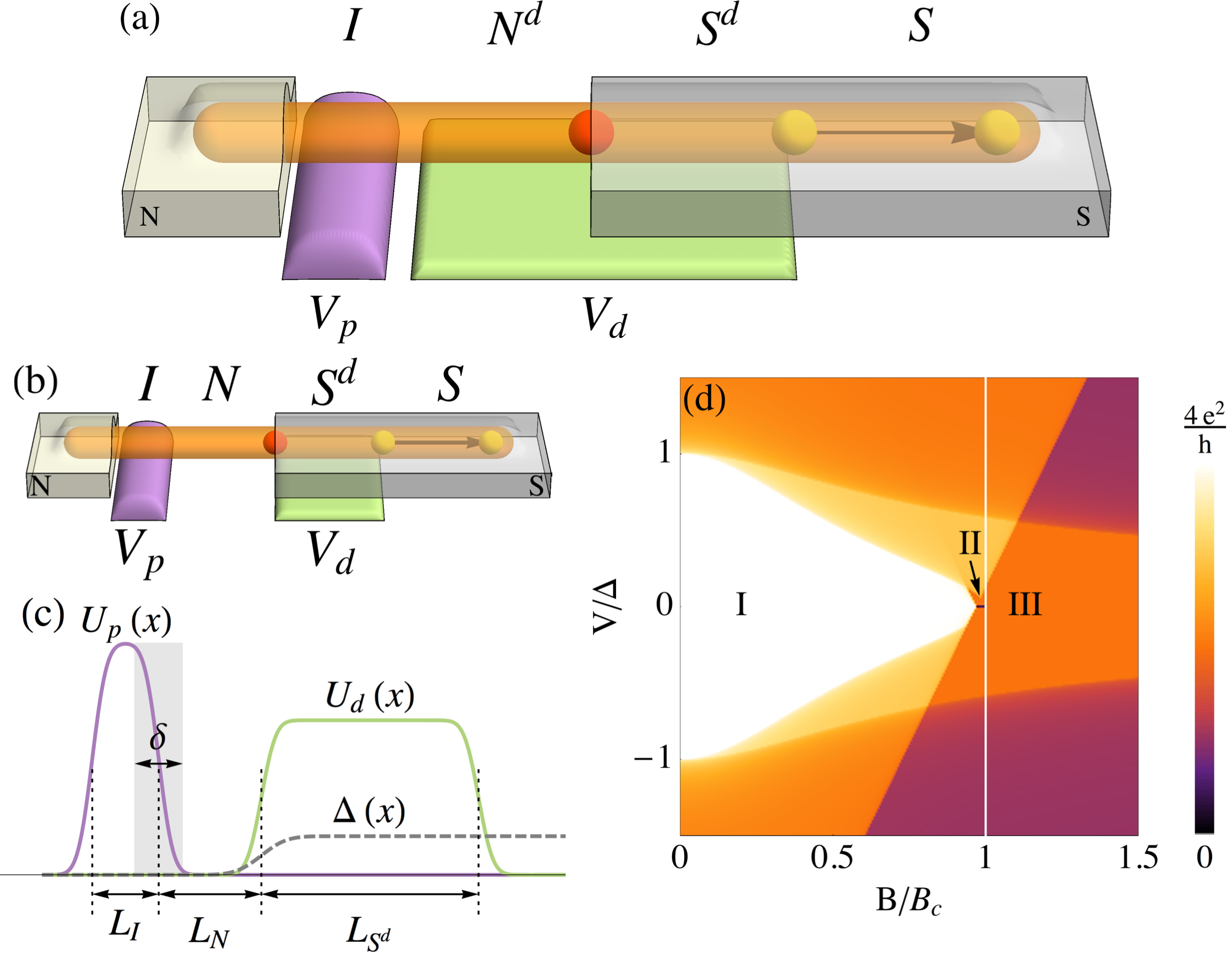} 
     \caption{(Color online) Schematics of the nanowire junction in the \ndsds\  (a) and \nsds\  (b) setups, and spatial variation of superconducting gap and potential profiles (c). Gate $V_d$ depletes the wire, while $V_p$ creates a tunnel contact (I) to the left (normal) reservoir. One (red) or two (red and yellow spheres) Majorana bound states may appear at the edges of the depleted region depending on the Zeeman field and gate voltage $V_d$. (d) Transport regimes for a transparent NS junction ($V_{d,p}=0$, $\mu=4\Delta$) in the Zeeman field - bias plane. 
}
   \label{fig:sketch}
\end{figure}

Motivated by this, we present here {a detailed study} of transport through normal-superconductor (NS) junctions containing topological wires. As in the experiment, the wires are tunnel-coupled to the normal reservoir by a pinch-off gate $V_p$ (to allow for transport spectroscopy using $dI/dV$), and are depleted by an additional gate $V_d$ in order to bring $B_c$ down to accessible fields (since the induced potential $U_d$ lowers the Fermi energy $\mu\rightarrow\mu-U_d$). The depletion profile, however, is necessarily \emph{inhomogeneous}, since it cannot extend deep into the superconducting side due to efficient screening, see Fig. \ref{fig:sketch} and Ref. \onlinecite{Delft-exp}. We find that for this class of devices, a number of distinct transport regimes arise as the various sections of the wire transition to different electronic phases. We characterize these regimes and the rich phenomenology 
that results beyond the simplest picture \cite{Sengupta:PRB01,Bolech:PRL07,Law:PRL09}. In particular, we address the question of whether the ZBAs are related to Majorana physics, and {confirm} that this is indeed the case for long depletion regions. 
We also analyze the development of ABSs close to zero energy when the pinch-off gate lies at a finite distance from the NS junction. 
Our main results are summarized in Fig. \ref{fig:exp}(e) where we demonstrate that the $dI/dV$ of realistic junctions with inhomogeneous depletion and multisubband filling  \emph{may not} show a distinct closing of the gap and yet exhibit ZBAs of Majorana origin. In most cases, these ZBAs show a residual splitting and {may} coexist with ABSs, features also observed in Ref. \onlinecite{Delft-exp}.

\sect{Model}
We first consider a one-dimensional NS junction [see Fig. \ref{fig:sketch}(a)], with a BCS-type Hamiltonian $H=H_0+H_\mathrm{pairing}$, where
\[
H_0=\int dx~\psi^\dagger(x)\left[\frac{-\partial_x^2}{2m}+i\alpha \sigma_y\partial_x+B\sigma_z+U(x)-\mu\right]\psi(x)
\]
and $H_\mathrm{pairing}=\int dx~ \psi^\dagger(x) i\Delta(x)\sigma_y\psi^\dagger(x)+\mathrm{h.c}$.

Here $\alpha$ is the SO coupling and $B$ is the Zeeman splitting (given by $B = g\mu_B\mathcal{B}/2$, where $\mathcal{B}$ is an in-plane magnetic field, $\mu_B$ is the Bohr magneton and $g$ is the nanowire g-factor).
We assume a position dependent pairing $\Delta(x)$ induced by the superconducting electrode such that $\Delta(x\to\infty)=\Delta$  and $\Delta(x\to-\infty)=0$. The term $U(x)=U_d(x)+U_p(x)$ is composed of two parts: $U_p(x)$ comes from the pinch-off gate $V_p$ in the normal region at a distance $L_N$ from the NS interface, and $U_d(x)$ models the potential induced by the depletion gate $V_d$ \footnote{This gate has turned out to be crucial for the observation of ZBAs in the Delft experiment.}. Gate $V_d$ may extend all the way into the normal side of the NS interface [case \ndsds, with a depleted length $L_d=L_{N^d}+L_{S^d}$, Fig. \ref{fig:sketch}(a)], or be limited to the end of the superconducting side [case \nsds, $L_d=L_{S^d}$, Fig. \ref{fig:sketch}(b)]. We will consider the former case first, where we cover different parametric regimes, and then turn to the second one, which is closer to the experimental setup \cite{Delft-exp}. Realistic experimental parameters are: $\Delta= 250 \mathrm{\mu eV}$ is the induced gap that, for an InSb effective mass $m=0.015m_e$, corresponds to a length scale $L_\Delta\equiv\hbar/\sqrt{m\Delta}\equiv 142 \mathrm{nm}$. Strong SO coupling, representative of InSb wires \cite{Nadj-Perge:PRL12}, is $\alpha=20$ meV nm, with SO length $L_{\rm SO}=\hbar^2/(m\alpha)=
200 \mathrm{nm}=1.4L_\Delta$ \footnote{These parameters, while probably being a good estimation, are not directly measured in the experiment but rather inferred from different samples in a different geometry. Thus significant changes in  $L_{SO}$ cannot be excluded.}.

\sect{Scales}
A localized MBS is formed at the boundary of a trivially gapped and a TS portion of the wire. At a point $x$ the wire will be in the TS phase if $\Delta(x)>0$ and 
\begin{equation}
B>\sqrt{[\mu-U(x)]^2+\Delta(x)^2}.
\end{equation} 
The asymptotic value of the critical field is the proper (bulk) critical field $B_c$. Apart from $B_c$, several other Zeeman scales dictate the junction's transport properties. The first one is the TS critical field in the depleted part of the superconducting wire, $B^d_{c}\equiv\sqrt{(\mu-U_d)^2+\Delta^2}$, that is smaller than $B_c$, as is the purpose of the depletion gate. It should be noted, however, that the depleted $S^d$ region has a \emph{finite} length, which crucially affects Majorana modes for $B^d_{c}<B<B_c$, as discussed later, while the $S$ portion is assumed infinite. Secondly, there is the field above
which the normal side of the wire becomes a helical liquid (momentum and spin become correlated). In the \nsds\   case (normal side not depleted), this is $B_h\equiv \mu$, which is typically slightly smaller than $B_c$, but bigger than both $B^d_{c}$ and the corresponding helical field in the \ndsds\  case, namely $B^d_{h}\equiv |\mu-U_d|<B^d_{c}$. Finally, there is the superconducting gap itself, $B_\Delta\equiv\Delta$, whose significance will become clear later.
All these scales ($B_\Delta$ plus $B^d_{c}<B_h<B_c$ in the \nsds\  case, or $B^d_{h}<B^d_{c}<B_c$ in the \ndsds), control different aspects of the junction's differential conductance in the $B$-$V$ plane.

\sect{Differential conductance}
The $dI/dV$ of a NS junction may be related to the intrinsic conductance at zero temperature by the expression \cite{Blonder:PRB82}:
\begin{equation}
\frac{dI(V)}{dV}=\frac{e^2}{h}\left[\mathcal{N}-\mathrm{Tr}(r_{ee}^\dagger r_{ee})+\mathrm{Tr}(r_{eh}^\dagger r_{eh})\right]_{\epsilon=V}.\nonumber
\end{equation}
Here, $\mathcal{N}$ is the number of propagating channels in the normal side at energy $\epsilon=V$, and $r_{ee}$ and $r_{eh}$ are their normal and Andreev reflection matrices.
These  matrices can be computed in a number of ways. The most flexible is the recursive Nambu Green's function approach, employed here (for full details, see \cite{Sup-Mat}).

\begin{figure} %  figure placement: here, top, bottom, or page
   \centering
   \includegraphics[width=0.85\columnwidth]{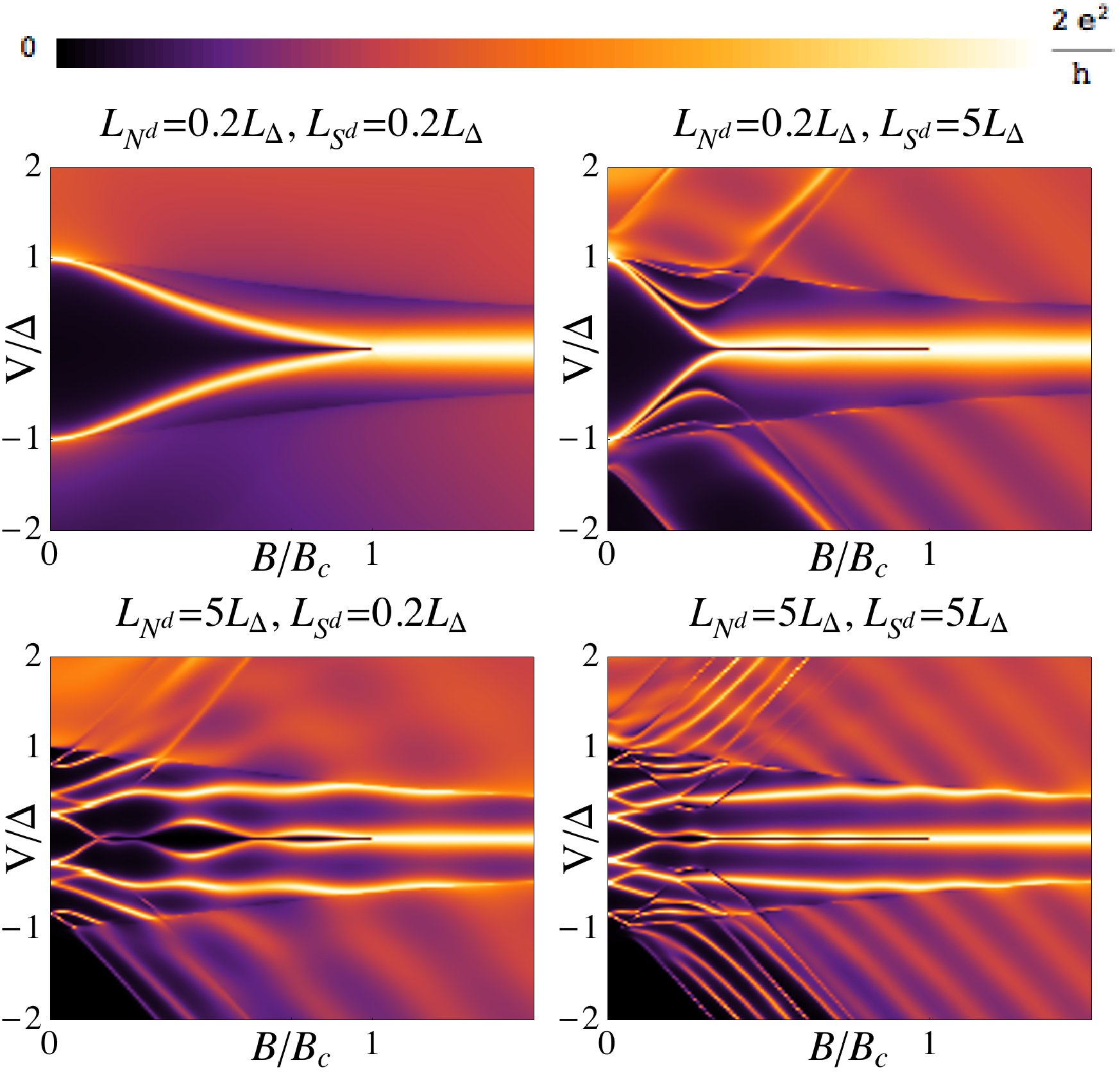} 
   \caption{(Color online) Density plots of the $dI/dV$ in the \ndsds\  junction ($\mu=4\Delta$, $U_d=3.25\Delta$, $U_p=25\Delta$, $\delta=0$) for $L_{\rm SO}=1.4L_\Delta$ as a function of bias voltage $V$ and Zeeman field $B$ with a tunnel pinch-off barrier and a depletion region of length $L_{N^d}+L_{S^d}$, Fig. \ref{fig:sketch}(a). Different columns feature increasing values of $L_{S^d}$ from left to right, whereas different rows feature increasing length $L_{N^d}$ from top to bottom.}
   \label{fig:dIdV}
\end{figure}

Before considering the effect of $U(x)$, we show the transport phase diagram [see Fig. \ref{fig:sketch}(c)] in the simple NS transparent limit, i.e. in a regime where the concepts of MBSs and ZBAs do no longer hold. We observe different transport regions in the $B$-$V$ plane characterized by an integer $dI/dV\approx n e^2/h$, with $n=0,1,2,3,4$. {Such} is the case of Cooper pair transport (region I, $n=4$) or single quasiparticle transport (region III, $n=2$). The latter is a TS regime, whose topology becomes evident in the $dI/dV$ despite the fact that the associated Majorana fermion is completely smeared out due to the gapless spectrum for $x<0$ \cite{Wimmer:NJOP11, Sau:PRL12, Gibertini:PRB12, Chevallier:PRB12}. Between these two regions, the helical regime is characterized by a fully suppressed zero bias conductance (region II, $n=0$). These results extend the concept of half-integer conductance quantization \cite{Wimmer:NJOP11} beyond linear response.

\begin{figure}[t] %  figure placement: here, top, bottom, or page
   \centering
   \includegraphics[width=\columnwidth]{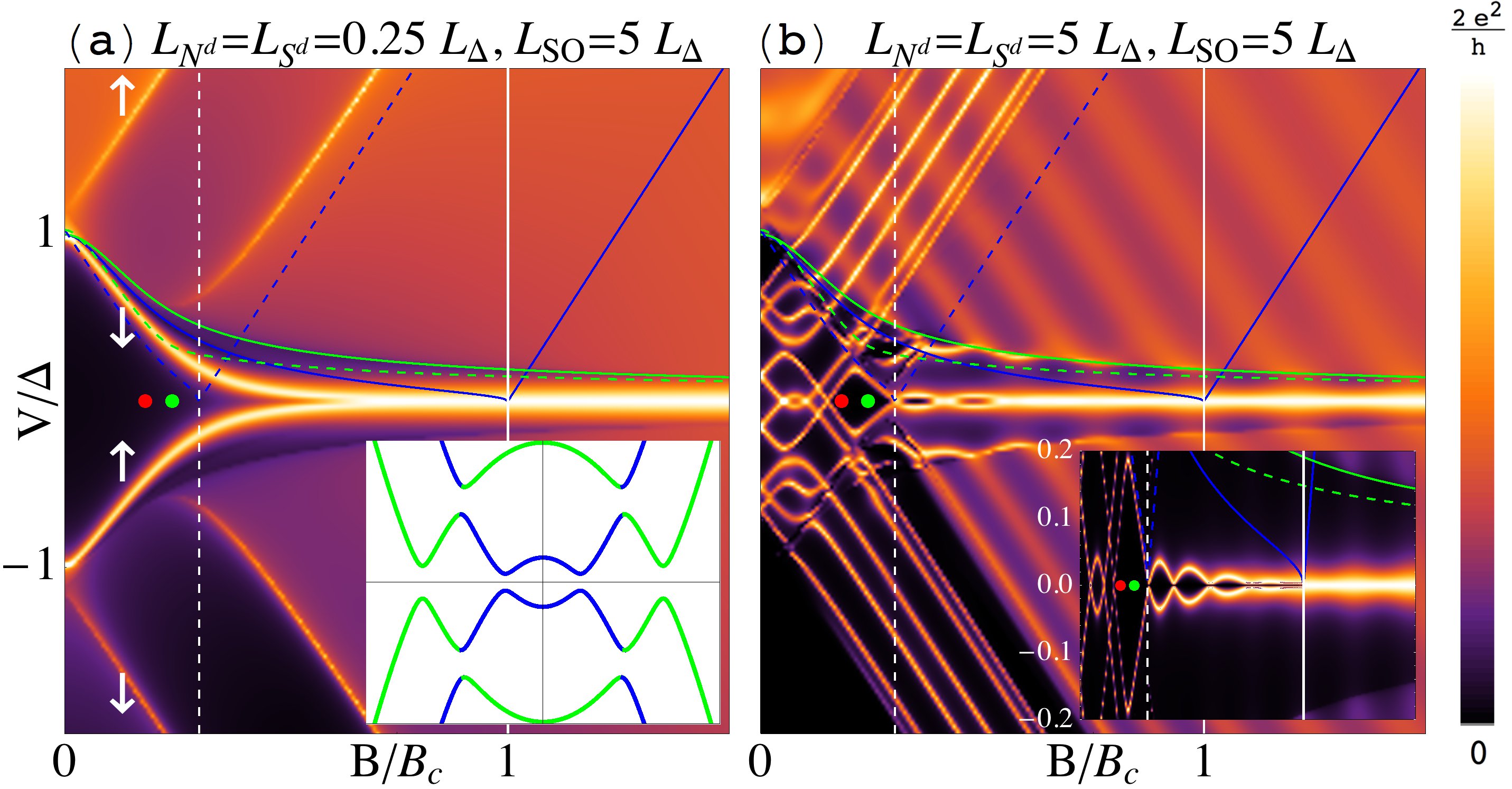} 
   \caption{(Color online) $dI/dV$ in a short (a) and long (b) \ndsds\ junction ($\mu=4\Delta$, $U_d=3.25\Delta$, $U_p=22\Delta$, $\delta=0$).  Coloured guidelines represent the gap at large (green) and small (blue) momentum (see bandstructure in the left inset) in the depleted (dashed) and bulk (solid) superconducting regions. Vertical solid (dashed) line marks $B_c$ ($B^d_c$). (a) For short depletion region, a mini gap transport regime is developed for $B>B_\Delta$ (green dot), unrelated to Majorana modes. (b) Long depletion region, however, allows for the existence of ABSs around zero energy below $B_h^d$ (red dot) and (split) Majorana fermions for $B<B_c$, see blowup in the right inset.  
   }\label{fig:shortlong}
\end{figure}

\begin{figure} %  figure placement: here, top, bottom, or page
   \centering
   \includegraphics[width=\columnwidth]{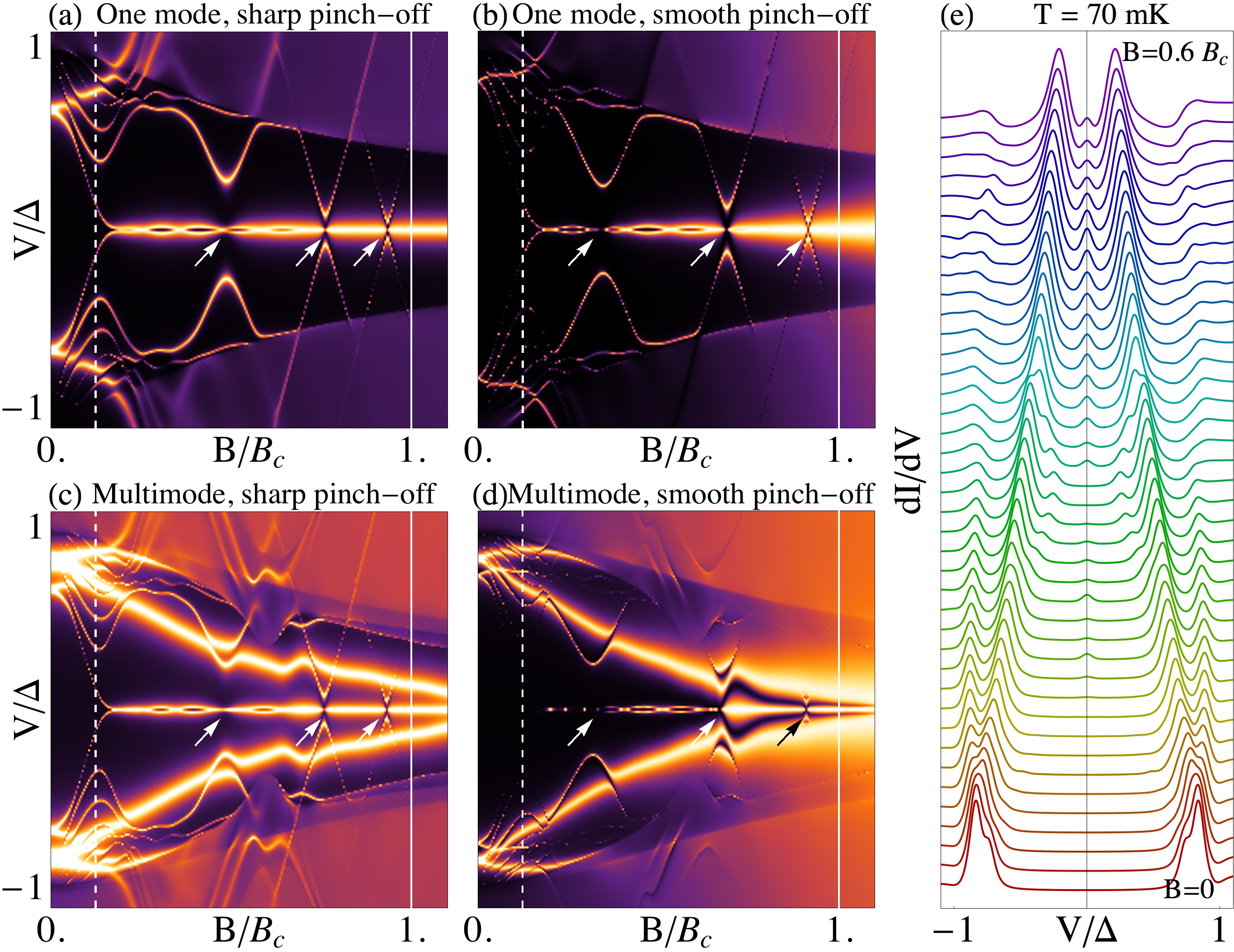} 
   \caption{(Color online) $dI/dV$ response at zero temperature of an \nsds\  junction similar to that of Ref. \onlinecite{Delft-exp}, for one open (spinful) mode (a,b) and two modes (c,d). ABSs coexist and suppress (white arrows) the ZBA from Majorana pairs within the depletion region. For increasing smoothness of the pinch-off profile $U_p(x)$ (b and d), the local gap edge peak is washed out, leaving an isolated ZBA at $B<B_c$ without a visible closing of a transport gap. Vertical guidelines and color scale like in Fig. \ref{fig:shortlong}. (e) Finite temperature dI/dV version of (d). Different curves are shifted for clarity from $B=0$ (bottom) to $B=0.6B_c$ in 40 steps. Parameters: $\mu=U_d=8\Delta$, $U_p\sim 20$-$30\Delta$, $L_N=450nm$, $L_S^d=1\mu$m, $\delta=100nm$, $B_c^d=0.12B_c$}\label{fig:exp}
\end{figure}

We now consider the \ndsds\  junction with the full $U(x)$. Its $dI/dV$ response (with $L_{\rm SO}=1.4L_\Delta$) is plotted in Fig. \ref{fig:dIdV}. Different panels cover different ratios $L_{\rm{N^d}}/L_\Delta$ and $L_{S^d}/L_\Delta$.
The tunnel barrier $U_p$ is tuned in each case to yield spectroscopic resolution in the transport response. A wide range of behaviours become apparent, that reflect the local density of states (DOS) at the pinch-off gate. The most paradigmatic one is probably the one in the top-left panel. It reflects the closing of the effective superconducting gap (marked by the gap-edge conductance peaks)  as $B$ increases. The gap-edge DOS peak transforms into a Majorana mode at zero energy for $B>B_c$ [pictured in Fig. \ref{fig:sketch}(a) as a red  sphere]. The gap reopens in the TS phase (see solid blue line in Fig. \ref{fig:shortlong}), but the local DOS at the contact is no longer peaked because the spectral weight is transferred to the Majorana mode. Hence, a prototypical three-pronged structure arises in the $B$-$V$ plane. However, the relevant phenomenology is by no means exhausted by this. Different scenarios arise at large  $L_{N^d}$ (Fig. \ref{fig:dIdV}, bottom row), with the development of ABSs, or large $L_{S^d}$ (right column), with the development of ZBAs below $B_c$. Further phenomenology is obtained by varying $L_{\rm SO}$ \cite{Sup-Mat}.

\sect{Short \ndsds\  junction}
We now analyze in more detail the short junction case, $L_{N^d}, L_{S^d}\ll L_\Delta$ (top-left panel in Fig. \ref{fig:dIdV}). The scale $B^d_{c}$ has little meaning in this case, since any pair of Majorana modes forming at the ends of a short TS wire will strongly hybridize into two conventional states with energies $\sim \pm\Delta$. The effect of increasing $L_{\rm SO}$ is to flatten the gap-edge $dI/dV$ peak (bright yellow) to lower energies, as shown in Fig. \ref{fig:shortlong}(a) ($L_{\rm SO}=5L_\Delta$).
At these large $L_{\rm SO}$, spin becomes a good quantum number and the Zeeman field splits the particle and hole bands of the S region into spin up/down subbands (white arrows in Fig. \ref{fig:shortlong}). Particle and hole gap edges anticross at zero energy for $B>B_\Delta$ (green dot), resulting in a minigap (which vanishes for $L_{\rm SO}\to\infty$) with almost flat edges near zero. This structure appears as a (split) ZBA below $B_c$ that is unrelated to Majorana mode formation \footnote{Note that this structure of the $dI/dV$ remains essentially unchanged in the absence of a depleted region.}.

\sect{Long \ndsds\  junction} 
Increasing the length $L_{N^d}$ of the depleted normal wire gives rise to ABSs in the $N^d$ region that are probed by the tunnel barrier $U_p$ (see bottom row of Fig. \ref{fig:dIdV}). Increasing $L_{\rm SO}$ we find these ABSs resonances approaching a degenerate zero energy crossing, see Fig. \ref{fig:shortlong}(b). These low energy ABSs, however, cease after the helical transition at $B^d_{h}$ (red dot). While the ABSs move up and down in energy for $B<B^d_{h}$ depending on their spin character, as soon as the $N^d$ region becomes helical, all ABSs disperse \emph{away} from zero energy with increasing $B$, since incoming states do not have spin partners anymore. Beyond the second threshold $B^d_{c}$ (vertical dashed white line), the local band gap closes in the $S^d$ region (dashed blue line) and a ZBA builds up, which is caused by Majorana fermion pairs forming in the $S^d$ region of length $L_{S^d}$, also assumed large. The finite $L_{S^d}$, however, produces a residual splitting of the two Majoranas, visible in the ZBA that oscillates with $B$ \cite{Cheng:PRL09,San-Jose:11a, Lim:12,Klinovaja:PRB12} as long as $B<B_c$. Above the bulk $B_c$, the rightmost Majorana escapes to $x\to \infty$ (see Fig. \ref{fig:sketch}), where it no longer overlaps with the one at the $N^dS^d$ interface, and the splitting vanishes [see inset of Fig. \ref{fig:shortlong}(b)].

\sect{\nsds\  junction} 
We now turn to the type of setup explored in Ref. \onlinecite{Delft-exp}, the \nsds\  case.  The crucial difference with the \ndsds\  setup is that the ABSs  {zero energy anticrossings} may coexist with {Majorana ZBAs in the} $S^d$ region, since the $N$ region becomes helical \emph{after} the $S^d$ region becomes topological. 
When that happens, {the zero energy ABSs repel the MBS wavefunction (usually delocalized into the $N$ region) back into the $S^d$ region, hence decoupling it from the lead. As a result,} the ZBA, as measured by the $dI/dV$, is suppressed (see arrows in Fig. \ref{fig:exp}, where $L_{\rm SO}=1.4L_{\Delta}$). In a single mode wire (Fig. \ref{fig:exp}, top row), this may happen only for $B<B_h<B_c$, since it requires an interface between a non-helical $N$ and a topological $S^d$ \footnote{Therefore, for single mode wires, ABSs cannot coexist with the ZBA near zero energy for $B>B_c$. Although this \emph{is} possible in multimode wires, the ZBA suppression does not occur in such case.}. Interestingly, the above behavior agrees with the experimental observation of an intermittent ZBA that disappears and reappears as an ABS anticrosses at zero energy. Constant-$B$ sweeps of $U_d$ are also found to closely correlate with the experiment \cite{Sup-Mat}.

A further feature apparent in the {experiments} is the absence of gap-edge singularities closing just before the formation of the ZBA. Up till now, all potential profiles have been assumed spatially abrupt [decay length $\delta=0$ in all profiles of Fig. \ref{fig:sketch}(c)]. Imperfect screening, however, will lead to gate-induced potentials that decay slowly along the wire, specially at low electron densities. When a smooth $U_p(x)$ profile is taken into account, the gap-edge peaks are quickly washed out, and the $dI/dV$ in the tunneling regime is no longer a perfect measure of the local density of states \footnote{This is in stark contrast with local probes such as the tunneling current from a metallic tip into the end of the nanowire, see e.g. Ref. \cite{Stanescu:PRB11}. Note moreover that a smooth $\Delta(x)$ leaves the $dI/dV$ largely unaffected, which allows us to bypass a self consistent calculation of the pairing profile.}. 
Unlike for sharp pinch-off barriers, transmission through smooth barriers mostly preserves longitudinal momentum. Since moreover barrier transmission is lower for smaller momenta, the gap-edge resonance closing around zero momentum at $B_c^d$ is poorly probed by a smooth barrier. Visibility is restored as $B$ increases beyond $B_c^d$, since then the Zeeman splitting increases the momentum components of all resonances, including the ZBA.
Fig. \ref{fig:exp}(b) shows this effect. Note on the other hand that the gap-edge signal at higher energies $\sim \Delta$ is roughly constant in $B$. This corresponds to the large-momentum band edge, 
not the true gap edge at small momentum.
The former band edge shows no sign of the different transitions (the TS transition at $B_c$ in particular), and never closes. Therefore, all zero energy structure appears disconnected from any closing of the gap, as measured by the $dI/dV$. 

Transport features connected to large momenta should be \emph{enhanced} in a multisubband systems. The pinch-off condition for tunneling spectroscopy requires a higher $U_p$ barrier in this case, to shut off the additional open modes, which will necessarily contribute with a stronger signal relative to the shallower, lower momentum mode. This is likely the experimental situation in Ref. \onlinecite{Delft-exp}, {where} the observed gap edge signal exceeds the amplitude of the ZBA.
In order to confirm this idea,  we have performed multisubband transport simulations (see implementation details in \cite{Sup-Mat}). In Fig. \ref{fig:exp}(c), (d), we show results corresponding to a fixed $\mu$ defined from the bottom of the topmost subband, which becomes topological at low magnetic fields, while the lower ones remain trivial. The single mode analysis carries over unchanged, albeit with a stronger gap edge and the associated high-momentum structure of the additional modes. Fig. \ref{fig:exp}(e) shows constant $B$ traces of Fig. \ref{fig:exp}(d) at finite temperature ($k_BT=70mK$). Thermal smearing washes out the ZBA splitting, whilst also reducing its height.

\sect{Conclusions}
We have identified various transport regimes in depleted NS nanowire junctions with SO coupling and Zeeman field. Depending on the Zeeman field, the wire as a whole may be in any given combination of helical/non-helical and trivial/topological phases for its normal and superconducting portions, respectively. These include a \emph{helical depletion} ($B_h^d<B<B_c^d$) and a \emph{topological depletion} ($B_c^d<B<B_h, B_c$) phases that arise at fields below those of the proper \emph{helical bulk} ($B_h<B<B_c$) and \emph{topological bulk} ($B>B_c$) regimes.  The different phases have distinct sub-gap signatures in transport, particularly if the depleted $S^d$ region is long enough. In this case, ZBAs appear that are caused by the formation of either a single (in the \emph{topological bulk} phase) or a pair (\emph{topologial depletion} phase) of Majorana modes in the junction. The latter is characterized by a residual ZBA splitting. In the case of short junctions, Majoranas cannot develop below $B_c$, although one may still distinguish between the \emph{conventional} and \emph{minigap} transport regimes (unrelated to Majorana physics) if $L_{\rm SO}$ is large. 

Apart from this general analysis, we have discussed a configuration similar to the one in the experiment of Ref. \onlinecite{Delft-exp}. We argue that at least a number of non-trivial features observed  in this experiment are consistent with most of our results corresponding to transport through a multi-mode non-helical normal/topological depleted superconductor/trivial bulk superconductor junction, hosting Majorana fermion \emph{pairs} within the central region [Fig. \ref{fig:exp}(d), (e)]. While a qualitative correspondence can be traced between our observations and the experiment, a more quantitative agreement is beyond the goal of our study. This is in part due to a considerable number of unknown parameters (precise $L_{\rm SO}$, screened gate potential profiles, localization and pair-breaking effects in the bulk superconductor \cite{Driessen:PRL12}, etc.). Moreover, other features that we find, such as the ZBA splitting \emph{oscillations} in $B$, have not been observed. Thus, a different physical origin of the measured ZBAs (reflectionless tunneling \cite{Wees:PRL92}, Kondo, etc) cannot be completely excluded. The latter, however, should be accompanied (at least in its most conventional form) with even-odd effects as a function of gate voltages \cite{Lee:PRL12}, which are not apparent in Ref. \onlinecite{Delft-exp}. Disorder (not considered here) might also be of relevance. However, its precise role on Majorana physics is currently under active investigation  \cite{Pientka:12,Bagrets-Altland:12,Liu:12}.

We are grateful to S. Frolov and L. Kouwenhoven for fruitful discussions. We acknowledge the support of the CSIC JAE-Doc program and the Spanish
Ministry of Science and Innovation through Grant No. FIS2008-00124/FIS (P.S.-J), FIS2009-08744 (E.P. and R.A.). This research was supported in part by the National Science Foundation under Grant No. NSF PHY05-51164.

\bibliography{biblio}

\begin{thebibliography}{40}
\bibitem [{Sup()}]{Sup-Mat}%
  \BibitemOpen
  \href@noop {} {}\bibinfo {note} {See Appendix for a description of the
  recursive Green's function algorithm for computing transport, including
  finite width and finite temperature. Also included is an extensive parametric
  study of differential conductance for varying geometric parameters and
  spin-orbit lengths, as well as differential conductance maps as a function of
  bias and depletion gate voltage.}\BibitemShut {Stop}%
\end{thebibliography}

\clearpage


\section*{Supplementary material}

# Supplementary Material for "Transport spectroscopy of NS nanowire junctions with Majorana fermions"

Elsa Prada[1], Pablo San-Jose[2], Ramón Aguado[1]
[1] *Instituto de Ciencia de Materiales de Madrid (ICMM-CSIC), Cantoblanco, 28049 Madrid, Spain*
[2] *Instituto de Estructura de la Materia (IEM-CSIC), Serrano 123, 28006 Madrid, Spain*
(Dated: September 20, 2012)

## I. NUMERICAL METHOD: RECURSIVE GREEN'S FUNCTION ALGORITHM

The Hamiltonian of the system with a single mode is $H = H_0 + H_{\mathrm{pairing}}$, where

$$H_0 = \int dx\, \psi^\dagger(x) \left[ \frac{-\partial_x^2}{2m} + i\alpha\sigma_y\partial_x + B\sigma_z + U(x) - \mu \right] \psi(x)$$

and $H_{\mathrm{pairing}} = \int dx\, \psi^\dagger(x) i\Delta(x)\sigma_y \psi^\dagger(x) + \mathrm{h.c}$ are defined on an unbounded one dimensional spatial continuum, times an SU(2) subspace for spin. A strategy to solve elastic transport is to discretize the continuum into an infinite tight-binding chain with lattice spacing $a$, where derivative $\partial_x \psi(x)$ is approximated by finite differences. The computed transport properties should be independent on the discretization parameter, and should therefore be constrained to low energies.

The way to deal with the pairing potential is to introduce Nambu space notation. Particle and holes are treated much like spin, as an extra quantum number. This is valid in the absence of additional interactions, and introduces a doubling of the Hilbert space. Wavefunctions and propagators are consequently doubled in size. Other than that, the recursive Green function algorithm to computing the scattering matrix[1] is the same than in systems without superconductivity. We summarize it in what follows.

The scattering matrix is computed by solving the Lippmann-Schwinger equation[1], that in the form most useful to us relates the full scattering states in the wire $|\Psi_\alpha\rangle$ at energy $\omega$ to the scattering states of the semi-infinite leads $|\Psi_\alpha^0\rangle$, the Green function $G(\omega)$ of the full wire, and the operator $V$ coupling the leads to the system.

$$|\Psi_\alpha\rangle = |\Psi_\alpha^0\rangle + (GV)_{\alpha\alpha'}|\Psi_{\alpha'}^0\rangle$$

The separation of the scattering problem into an easy *lead* subproblem, coupled to a complicated *system* is common. It requires identifying a repeating block in the Hamiltonian at $x \to \pm\infty$ (a unit cell), each of which has matrix elements $h$ under $H$, and are coupled to adjacent blocks by matrix elements $v$. Everything else falls into the *system* portion, and transport through it is computed recursively with the boundary conditions imposed by the leads.

To solve the leads, one assumes they are semiinfinite and decoupled from the system, and uses any method available to solve their Green function $g$ for a given $h$ and $v$, evaluated on the last unit cell, the one coupled to the system. This can be done by self-consistent iteration (imposing that adding a new unit cell to a semi-infinite lead does not change the surface Green function $g$), or by turning the problem into a generalized eigenvalue problem. In any case, $g$ should be obtained under the boundary condition that it be non-divergent as one advances far into the leads (it should be the *retarded* $g$).

The next step is to slice the discretized system portion into subparts analogous to the unit cells of the leads. Unlike for the leads, these pieces will not be identical, and there will be a finite number of them. They should, however, be coupled only to nearest neighboring slices. It is always possible to slice the system in this way. To return to the scattering matrix, one needs the full $G$ of the leads+system problem. This is obtained by including in the system Hamiltonian the self-energy $\Sigma = VgV^+$ induced on the surface slices of the system by the two (or more) leads, which reduces the problem to the system alone (leads are incorporated exactly through $\Sigma$). This suggests a good way to define 'slice zero' of the system, i.e. the surface slice coupled to the leads.

Once the leads are integrated out, the full propagator $G_{00}$ within the surface slice of the system is computed recursively. Step zero is a simple matrix inversion

$$G_{00}^{(0)} = (\omega - H_{00} - \Sigma)^{-1},$$

where $H_{nm}$ are the matrix elements of $H$ between slice $n$ and $m$. Likewise $G_{nm}^{(k)}$ is the propagator between slice $n$ and $m$ including the $k$ first slices. The next step is to add the next slice through the Dyson equation, $G_{00}^{(1)} = G_{00}^{(0)} + G_{00}^{(0)} H_{01}(\omega - H_{11})^{-1} H_{10} G_{00}^{(0)}$. The term $H(\omega - H)^{-1} H$ is the self energy brought in by the additional slice. This can be repeated until all slices are exhausted. The general recursive relation is

$$G_{00}^{(k)} = G_{00}^{(k-1)} + G_{0,k-1}^{(k-1)} H_{k-1,k} (\omega - H_{kk})^{-1} H_{k,k-1} G_{k-1,0}^{(k-1)}$$

The recursion requires to keep track of $G_{00}$, $G_{0k}$ and $G_{k0}$ in each step. The computational cost scales linearly in the number of recursion steps (the number of slices) and like the cube of the typical slice size, which should include spin and Nambu space dimensions in our case. Therefore, in a purely single mode wire, all slices have size 4, while $M$ mode wires will have slice sizes $4M$.

Once obtained $G_{00}$, it can be plugged into the $G$ in the Lippmann-Schwinger equation, since the coupling $V$ connects the leads only to slice zero, by definition. The

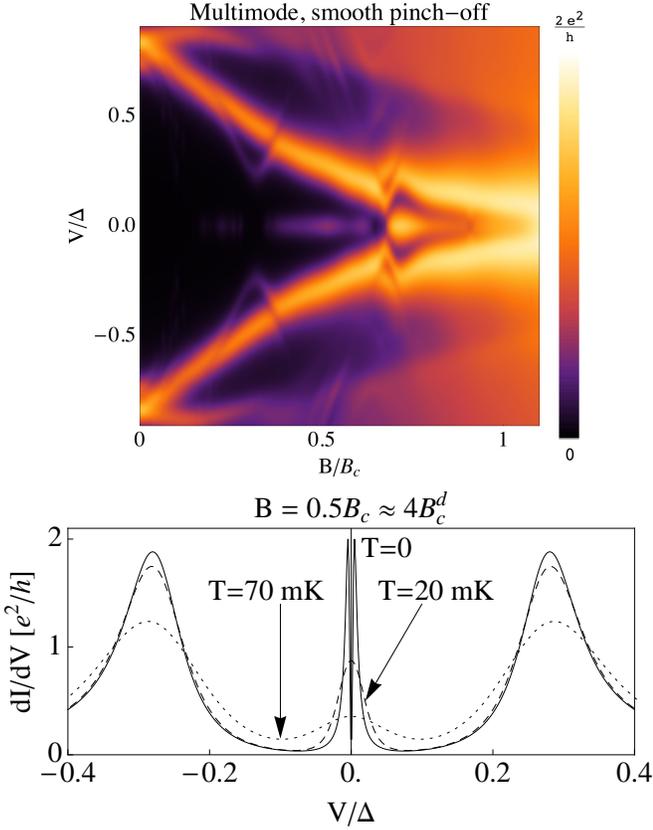

FIG. S1. Thermal smearing (top) at $T = 70$mK of the $dI/dV$ response in Fig. 4(d), with a section at fixed $B = 0.5B_c$ for increasing temperatures (bottom). Note the suppression of the ZBA amplitude far below the quantized zero temperature value $2e^2/h$.

scattering matrix can then be read off by decomposing $|\Psi_\alpha\rangle$ into right-propagating and left-propagating modes in each lead. In our simple case, this can be done by computing the bandstructure of the lead, but in a more general case the decomposition may be performed by analyzing $g$. Additionally, to compute the differential conductance $dI/dV$ in an $\mathcal{N}$ channel Normal-Superconductor junction

$$\frac{dI(V)}{dV} = \frac{e^2}{h}\left[\mathcal{N} - \mathrm{Tr}(r_{ee}^\dagger r_{ee}) + \mathrm{Tr}(r_{eh}^\dagger r_{eh})\right]_{\epsilon=V}$$

a decomposition into the electron ($e$) and hole ($h$) sectors of the reflection matrix has to be carried out.

## II. GENERALIZATION TO A MULTISUBBAND NANOWIRE

Applying the above recursion technique to multisubband nanowires is relatively straightforward. It is essentially achieved by considering a finite width in the transverse ($y$) direction of the wire. The $y$ dimension is then discretized much like the $x$ direction, and included into the mesh to solve recursively. Some physical details must be observed, however. The spin-orbit coupling in this case should include a $\partial_y$ component. For Rashba coupling, for example, this introduces an additional $\int dx\ \psi^\dagger(x)(-i\alpha\sigma_x\partial_y)\psi(x)$ term in $H$. Additionally, one must consider the energy shift of each of the resulting bands in the leads due to confinement energy $\int dx\ \psi^\dagger(x)\left(\frac{-\partial_y^2}{2m}\right)\psi(x)$. The lead integration and recursion procedure is carried out in the same fashion, albeit with larger sized slices and more costly matrix inversions.

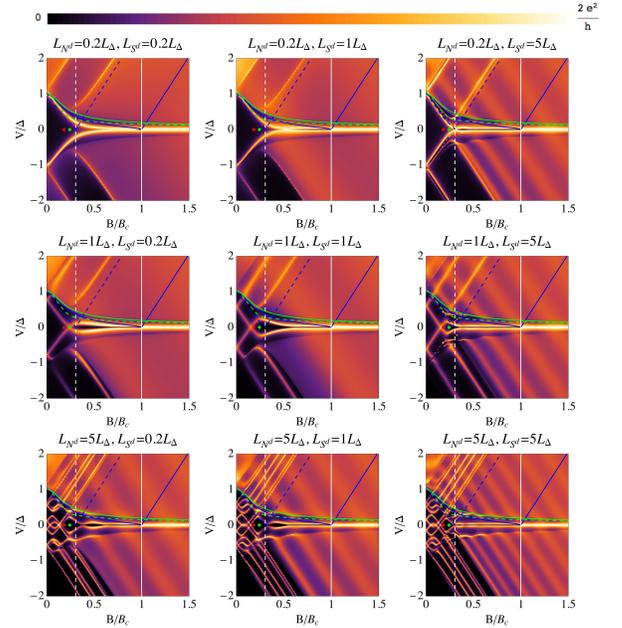

FIG. S2. $dI/dV$ in a $N^d S^d S$ junction for various lengths of the superconducting depleted region, $L_{S^d}$, and the normal depleted region, $L_{N^d}$, and weak spin-orbit coupling, $L_{\mathrm{SO}} = 5L_\Delta$. Different columns feature increasing values of $L_{S^d}$ from left to right, whereas different rows feature increasing length $L_{N^d}$ from top to bottom.

## III. EFFECT OF FINITE TEMPERATURE

Finite temperatures can be incorporated into the simulation rather easily, since its effect on $dI/dV|_{k_B T}$ amounts to a convolution of the zero temperature result $dI/dV|_0$ and the derivative of the Fermi-Dirac distribution:

$$\left.\frac{dI}{dV}\right|_{k_B T} = \int_{-\infty}^{\infty} dV\ \left.\frac{dI}{dV}\right|_0 \frac{d}{dV}\frac{-1}{1 + e^{V/k_B T}}. \quad \mathrm{(S1)}$$

In the experiment[2], a typical temperature is $T \sim 70$mK, which is a rather small scale in units of the superconducting gap $\Delta\ (= 250\mu eV)$, $K_B T/\Delta \approx 0.024$. However, as seen in Fig. S1, such smearing has a strong effect in reducing the height of the narrow resonances at zero temperature. The width of the resonances in units of

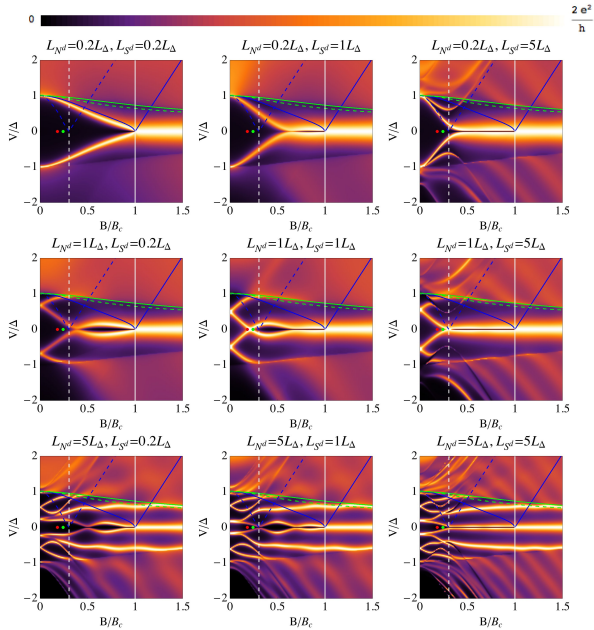

FIG. S3. $dI/dV$ in a $N^dS^dS$ junction for various lengths $L_{S^d}$ and $L_{N^d}$ and intermedium spin-orbit coupling, $L_{\rm SO} = 1L_\Delta$.

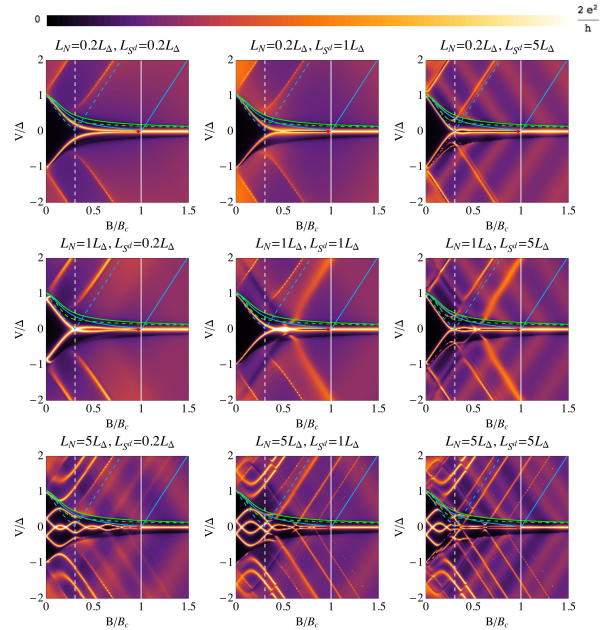

FIG. S5. $dI/dV$ in a $NS^dS$ junction for various lengths $L_{S^d}$ and $L_{N^d}$ and weak spin-orbit coupling, $L_{\rm SO} = 5L_\Delta$.

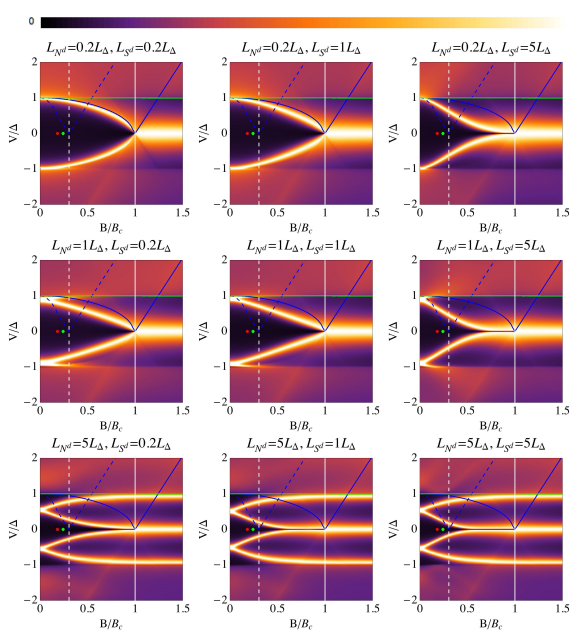

FIG. S4. $dI/dV$ in a $N^dS^dS$ junction for various lengths $L_{S^d}$ and $L_{N^d}$ and very strong spin-orbit coupling, $L_{\rm SO} = 0.2L_\Delta$.

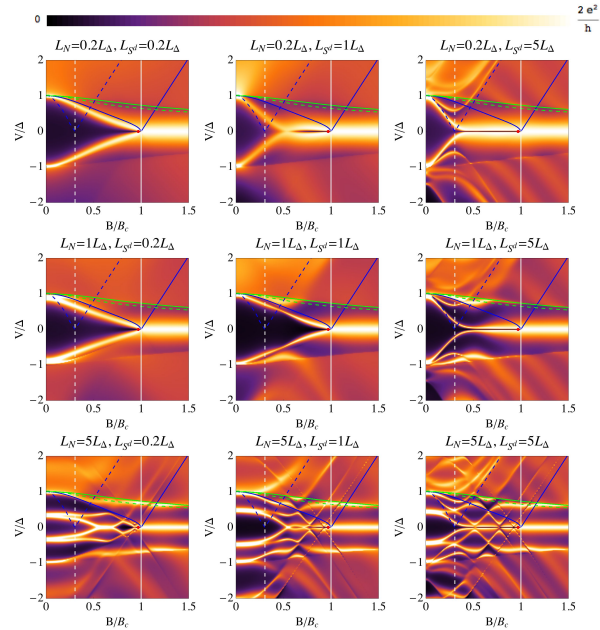

FIG. S6. $dI/dV$ in a $NS^dS$ junction for various lengths $L_{S^d}$ and $L_{N^d}$ and intermedium spin-orbit coupling, $L_{\rm SO} = 1L_\Delta$.

## IV. EXPLORING THE PARAMETER SPACE IN THE $dI/dV$ RESPONSE

The different characteristics exhibited in transport spectroscopy for spin-orbit lengths of 200nm, as corresponds to InSb wires, are presented in the main text in Fig. 2. Here we extend this analysis (also for spatially abrupt potentials) by considering wires with weaker

$k_BT$ gives a sense of the thermal suppression of the signal. Since, moreover, in multimode wires the width of the zero-bias anomaly (ZBA) coming from Majorana modes is strongly suppressed by the pinch-off of deeper modes, the overall ZBA amplitude at finite temperatures appears to be much smaller than the nominal zero-temperature $2e^2/h$.

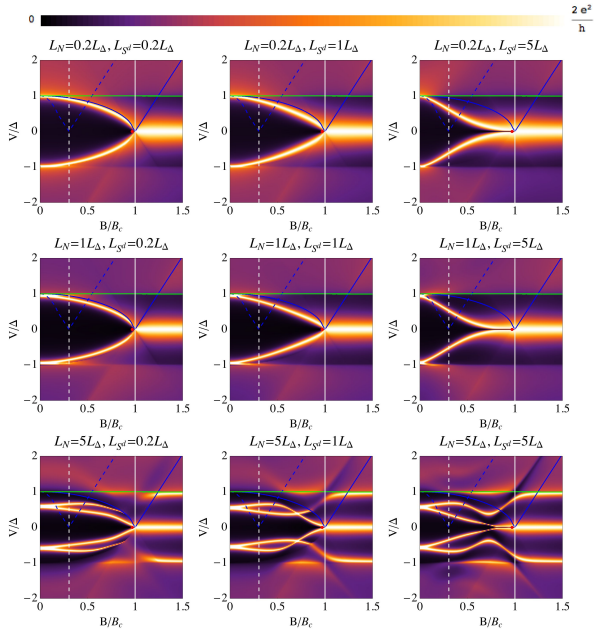

FIG. S7. $dI/dV$ in a $NS^dS$ junction for various lengths $L_{S^d}$ and $L_{N^d}$ and very strong spin-orbit coupling, $L_{SO} = 0.2L_\Delta$.

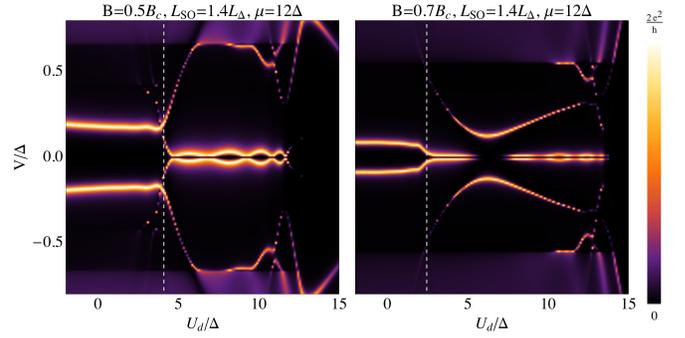

FIG. S8. Zero temperature $dI/dV$ at constant Zeeman field $B$ below $B_c$, $B = 0.5B_c$ in (a) and $B = 0.7B_c$ in (b). The value of $B_c^d = \sqrt{(\mu - U_d)^2 + \Delta^2}$ varies as the depletion potential $U_d$ is swept, and a transition occurs at $B_c^d = B$, dashed white line. When $B > B_c^d$, the depleted $S^d$ region enters a topological superconducting phase, and Majorana bound states are formed. The residual splitting is much smaller than the resonance spacing of the Majorana mode precursors visible at $B < B_c^d$.

and stronger spin-orbit coupling. Figs. S2-S4 present the results for $N^dS^dS$ wires, while Figs. S5-S7 correspond to $NS^dS$ wires. Note that, in all these panels, there is a certain asymmetry between the $+V$ and the $-V$ behaviour of the $dI/dV$ response[3]. This is due to the asymmetric electron-hole quasiparticle transport behaviour that is present in systems where the Andreev approximation doesn't hold, i.e., in systems in which $\Delta$ is not much smaller than $\mu$. This is the case of the kind of semiconducting nanowires that we study here, where we use $\mu = 4\Delta$ in all these plots (other parameters are: $U_d = 3.25\Delta$, $\delta = 0$ as corresponds to sharp potential profiles, and $U_p \sim 30\text{-}70\Delta$). Note that this effect is also present in the $dI/dV$ calculations of the main text.

The following general features are apparent as spin orbit strength is increased (spin-orbit length $L_{SO}$ is reduced): (i) the spacing and the width of Andreev bound states (ABSs) increase; (ii) the band edge at large momentum (green lines) becomes less sensitive to Zeeman coupling $B$ (spin becomes more polarized in plane by the spin-orbit coupling); (iii) Although the formation of Majorana modes and their corresponding ZBAs are not affected by the strength of the spin-orbit coupling for $B > B_c$, the MBS pairs forming in the depleted region for $B_c^d < B < B_c$ exhibit a splitting due to wavefunction overlap that grows with spin-orbit coupling strength. On the other hand, a weak spin-orbit coupling has the disadvantage that the excitation gap above the Majorana modes is diminished, which has important implications for their coherent manipulation for quantum computation purposes. If $L_{SO}$ is smaller than $L_\Delta$, and the depleted region is long enough so as to suppress the overlap splitting, one could achieve quasidegenerate Majorana pairs isolated from the quasiparticle continuum by a gap equal to a full $\Delta$, improving the chances to perform coherent non-Abelian braiding transformations of the wire's ground state at $B_c^d < B < B_c$.

Note once more that zero energy ABS anticrossings may coincide with Majorana pairs in the depleted region (above $B_c^d$) in the $NS^dS$ junction case, bottom rows of Figs. S5-S7. Such circumstance is not possible in the $N^dS^dS$ junctions of Figs. S2-S4. For long enough $L_{S^d}$ lengths (bottom-right panels), the MBSs around zero energy decouple from the pinch-off probe, and the ZBA signal is suppressed. This wavefunction repulsion effect between ABSs and MBSs as seen in the experiments, only arises between resonances belonging to the same subband, at least in the absence of subband mixing effects such as disorder.

## V. $dI/dV$ RESPONSE UNDER A SWEEP OF THE DEPLETION GATE

Experiment in Ref. 2 includes some results on the $dI/dV$ response in $NS^dS$ junctions as the depletion gate $V_d$ is swept at constant Zeeman field $B$ (see e.g. Fig. S5 in their supplementary material, $V_d$ is called $V_1$ there). The traces revealed what appears to be a split Majorana ZBA at small values of $V_d$. The authors point out this splitting *could* be related to overlap of Majorana bound states, but that more work needs to be done to confirm this interpretation. Our simulation, see Fig. S8, shows that the type of splittings obtained in the experiment are probably not due to Majorana overlap, but rather to a transition from $B_c^d > B$ (to the left of the dashed white line) to $B_c^d < B$ (to the right of the dashed white line) as $U_d$ is swept. Hence, the split ZBA is actually the remnants of the gap closing that is to be expected

in the standard picture of Majorana state formation, in the form of a low-lying ABS that is the precursor of a Majorana pair in the depleted zone. The actual splitting due to Majorana overlap occurs when the split resonance merges into a much narrower ZBA, whose residual splitting cannot be resolved experimentally at the working temperatures.


%%%%%%%%%%%%%%%%%%%%%%%%%%%%%%%%%%%%%%%%%%%%%%%%%%

\end{document}